\documentclass[twocolumn,prl,superscriptaddress,showpacs,showkeys]{revtex4}
\usepackage{graphicx}
\usepackage{graphics}
\usepackage{amsmath}
\usepackage{amssymb}
\usepackage{amsfonts}

\begin{document}

   \title{Flow of foam past an elliptical obstacle}

\author{Benjamin Dollet}
\email{b.dollet@utwente.nl} \altaffiliation{Present address:
Physics of Fluids, University of Twente, PO Box 217, 7500AE
Enschede, The Netherlands.} \affiliation{Laboratoire de
Spectrom\'etrie Physique, BP 87, 38402 Saint-Martin-d'H\`eres
Cedex, France} \altaffiliation{UMR 5588 CNRS and Universit\'e
Joseph Fourier.}
\author{M\'{e}lanie Durth}
\affiliation{Laboratoire de Spectrom\'etrie Physique, BP 87, 38402
Saint-Martin-d'H\`eres Cedex, France} \altaffiliation{UMR 5588 CNRS and
Universit\'e Joseph Fourier.}
\author{Fran\c cois Graner}
    \affiliation{Laboratoire de
Spectrom\'etrie Physique, BP 87, 38402 Saint-Martin-d'H\`eres Cedex, France}
\altaffiliation{UMR 5588 CNRS and Universit\'e Joseph Fourier}

\date{\today}

\begin{abstract}

To investigate the link between discrete, small-scale and
continuous, large scale mechanical properties of a foam, we
observe its two-dimensional flow in a channel, around an
elliptical obstacle. We measure the drag, lift and torque acting
on the ellipse {\it versus} the angle between its major axis and
the flow direction. The drag increases with the spanwise
dimension, in marked contrast with a square obstacle. The lift
passes through a smooth extremum at an angle close to, but smaller
than 45$^\circ$. The torque peaks at a significantly smaller
angle, 26$^\circ$. No existing model can reproduce the observed
viscous, elastic, plastic behavior. We propose a microscopic
visco-elasto-plastic model which agrees qualitatively with the
data.

\end{abstract}



\maketitle


A foam is a model to study  materials which are viscous, elastic,
and plastic. This complex, ubiquitous behavior is exploited in
numerous applications, such as ore separation, drilling and
extraction of oil, food or cosmetic industry \cite{Weaire1999},
but is not yet fully understood \cite{Hohler2005}. Foam rheology
is thus an active research area, and recent studies provide
insight to understand the interplay between the bubble scale and
the whole foam behavior \cite{Denkov2005,Cantat2005} and to unify
elasticity, plasticity and viscosity \cite{Weaire}. Here, we study
the flow of foam around an ellipse, where the measured lift, drag
and torque show the whole complexity of foam rheology, which
strongly constrains possible models: simple ones do not capture
the observed features. We propose an elastoplastic model which
describes well the data. We discuss the generality, implications
and limitations of this model.

We have built a foam channel \cite{Dollet2005} to investigate a 2D
steady flow and measure the force it exerts on an obstacle (Stokes
experiment
\cite{Dollet2005,Courty2003,deBruyn2004,Alonso2000,Cantat2005b}).
Briefly, a 1 m long, 10 cm wide tank is filled with deionized
water with 1\% of commercial dish-washing liquid (Taci, Henkel).
Its surface tension is $\gamma = 26.1\pm 0.2$ mN m$^{-1}$, and its
kinematic viscosity is $\nu = 1.06\pm 0.04$ mm s$^{-2}$. Several
computer controlled injectors blow nitrogen  in the solution to
form a horizontal monolayer of bubbles of average thickness $h =
3.5$ mm, confined between the bulk solution and a glass top plate
(quasi-2D foam) \cite{Cox2003}. This foam is monodisperse (bubble
area at channel entrance: $A_0 = 16.0\pm 0.5$ mm$^2$) and its
fluid fraction is estimated to be around 7\% \cite{Raufaste}.

\begin{figure}
\includegraphics[width=8cm]{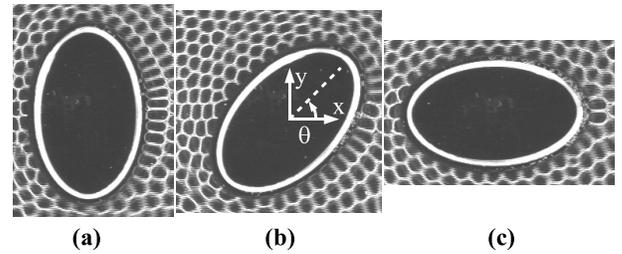}
\caption{\label{Ellipses} Top view of the elliptical obstacle and of the
surrounding bubbles. Ellipse (a) perpendicular to the flow, (b) tilted in the
flow, (c) parallel to the flow. The $x$ axis is the direction of the
flow direction and of positive drag,   $y$   is the direction of
positive lift,
and $\theta$ (between $0$ and 90$^\circ$, by symmetry) is the angle
between $x$ and the major axis of
the ellipse.}
\end{figure}

We study here  the simplest shape which symmetry is low enough to
observe simultaneously drag, lift and torque: the ellipse  (Fig.
\ref{Ellipses}). The obstacle has a major axis $2a= 48$ mm and
minor axis $2b= 30$ mm. It floats freely just below the top glass
surface, without solid friction. The upper end of an elastic fiber
passes through a hole in the bottom of the obstacle, ensuring its
free rotation, while its lower end is fixed, so that  a top view
of the obstacle displacement from its position at rest measures,
with a precision better than 0.1 mN, the force exerted by the foam
on the obstacle.

\begin{figure}
\includegraphics[width=8cm]{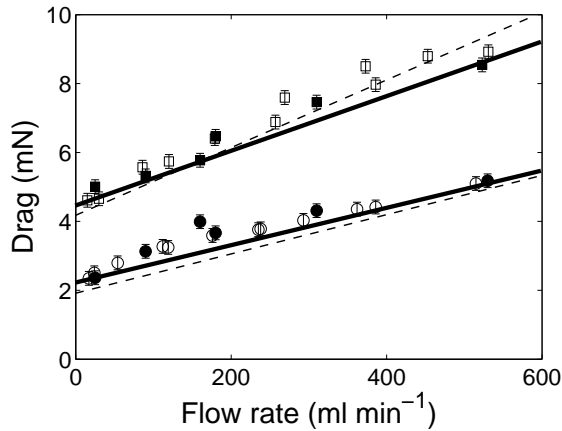}
\caption{\label{F(Q)} Drag exerted by the flowing foam on an elliptical
obstacle, \emph{versus} the flow rate: $\bullet$, $\theta=0^\circ$ (Fig.
\protect\ref{Ellipses}a); $\blacksquare$, $\theta=90^\circ$ (Fig.
\protect\ref{Ellipses}c). Bold lines are linear fits to the data. Data (open
symbols) and fits (dashed lines) for circular obstacles
   ($\circ$, 30 mm; $\square$, 48 mm diameter) from
Ref. \protect\cite{Dollet2005} are plotted for comparison. }
\end{figure}

\begin{figure}[b]
\includegraphics[width=8cm]{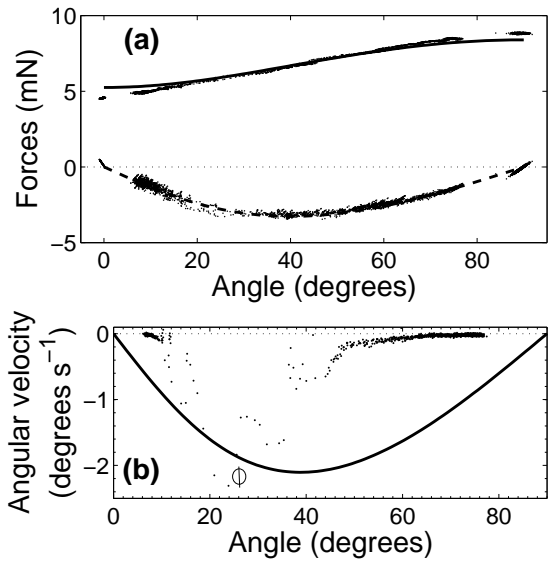}
\caption{\label{ForcesCoupleExpTh} (a) Drag (positive values) and
lift (negative values), (b) angular velocity $\dot{\theta}$
\emph{versus} $\theta$. Data correspond to the time-dependent
experiments with four initial orientations: 76 (same data as in
Fig. \protect\ref{Angle,F(t)}), 64, 48 and 18$^\circ$. Also shown
for lift and drag are data from the steady orientations,
$\theta=0^\circ$ and 90$^\circ$. In (a), the plain curve is the
fit by the spanwise dimension (\ref{LeadingLength}), and the
dashed one is the fit by Eq. (\ref{CoupleIntegral}) for the lift;
both fits are up to a free prefactor. In (b), data for $15^\circ <
\theta < 40^\circ$  are noisy  but compatible with the insert of
Fig. \ref{Angle,F(t)} (open circle). The curve is the fit by Eq.
(\ref{CoupleIntegral}), in arbitrary units.
}
\end{figure}

We measure the drag  in the parallel orientation ($\theta=0$),
which is stable (see below); and  in the perpendicular one
($\theta=90^\circ$), which is unstable (but where the ellipse can
remain for one hour, enough to perform steady flow measurements).
The results (Fig. \ref{F(Q)}) are very close to that for circles
of diameters 30 and 48 mm, respectively: this suggests (see also
Fig. \ref{ForcesCoupleExpTh}a) that drag is proportional to the
spanwise direction (along the $y$ axis) $\ell$ of the ellipse:
\begin{equation}\label{LeadingLength}
\ell = 2\sqrt{a^2 \sin^2 \theta + b^2 \cos^2 \theta} .
\end{equation}

  In a steady flow (530 ml min$^{-1}$,
{\it i.e. } a velocity of 2.5 cm s$^{-1}$), we start from a given
initial orientation (76, 64, 48 or 18$^\circ$), let the ellipse
rotate freely  to its parallel stable orientation, and measure the
angle, drag and lift  (Fig. \ref{Angle,F(t)}). The angular
velocity strongly increases in the range $15^\circ < \theta <
40^\circ$ (with a peak at 26$^\circ$) and does not depend on the
initial orientation (insert of Fig. \ref{Angle,F(t)}). Moreover,
the forces correlate to $\theta$; we thus eliminate the time and
plot
   the dependence of drag and lift with $\theta$
   (Fig. \ref{ForcesCoupleExpTh}a).
All the forces data collapse on two master curves, one for the
drag and one for the lift. The drag increases roughly linearly
with $\theta$ except very close to 0$^\circ$ and 90$^\circ$, where
it is extremal by symmetry (it equals 4.5 mN for 0$^\circ$ and 8.8
mN for 90$^\circ$). As suggested before, the experimental angular
dependence of the drag is close to the one of the spanwise
dimension, despite small discrepancies for angles close to
0$^\circ$ and 90$^\circ$. The lift vanishes at 0$^\circ$ and
90$^\circ$, as expected by symmetry; it is negative (downwards)
for angles between 0$^\circ$ and 90$^\circ$, with a maximal value
of 3 mN at an angle of about 40$^\circ$.

\begin{figure}
\vspace{1cm}
\includegraphics[width=8cm]{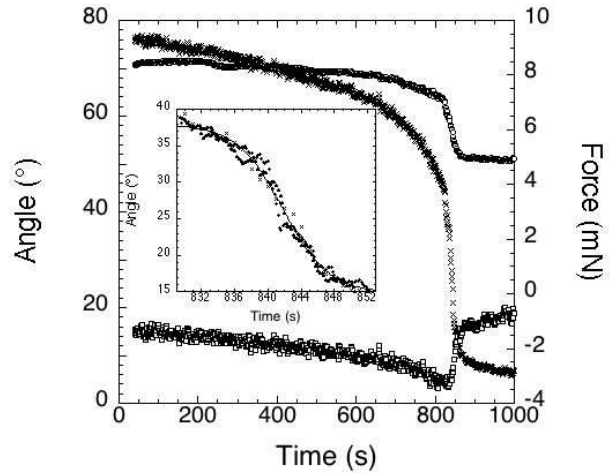}
\caption{\label{Angle,F(t)} Angle ($\times$), drag ($\circ$) and
lift ($\square$) of the ellipse, \emph{versus} time, for an
initial angle of 76$^\circ$. Insert: zoom on the region of quick
variation of the angle. The data for three different initial
orientations ($\times$: 76$^\circ$, $+$: 64$^\circ$,
$\vartriangle$: 48$^\circ$) are superimposed, by translating the
time axis. The solid line is a fit to all  data with a hyperbolic
tangent profile, indicating that the maximum angular velocity is
$-2.1\pm 0.1^\circ$ s$^{-1}$ for an angle of $26.4\pm 0.1^\circ$.}
\end{figure}

These measurements are independent of the initial orientation,
even in the region of quickest variation (insert of Fig.
\ref{Angle,F(t)}). This suggests that the results, obtained in
transient regimes, would be the same if we could fix $\theta$ to
perform  steady flow  measurements. In fact, at a lower flow rate
(25 ml min$^{-1}$), we observe very similar tendencies, although
more noisy (data not shown). It is thus natural to neglect the
obstacle's inertia, and assume that the torque exerted by the
flowing foam is exactly balanced by a friction torque (arising
mainly from viscous dissipation in the capillary bridge between
the ellipse and the top plate). Furthermore, the angular velocity
$|\dot{\theta}|$ is lower or comparable to 1$^\circ$ s$^{-1}$
(Fig. \ref{ForcesCoupleExpTh}b), hence the associated Reynolds
number $a^2 |\dot{\theta}|/\nu$ does not exceed 10. We can thus
assume that the friction torque
  is
proportional to  the angular velocity $\dot{\theta}$, then Fig.
\ref{ForcesCoupleExpTh}b represents (up to
  an unknown multiplicative constant characterizing the dissipation)
   the torque exerted by the foam.
It displays a peak around 26$^\circ$, compatible with   Fig.
\ref{Angle,F(t)}.

%

Fig. \ref{ForcesCoupleExpTh}b shows that the torque is negative
for all positive $\theta$ values. Thus  the only stable
orientation of the ellipse is the parallel one, $\theta=0^\circ$.
   This contrasts with the Newtonian case, where long objects
settle broadside-on \cite{Feng1994}. Note that in the case of a
Stokes flow (without inertia nor elasticity), every orientation of
the ellipse would be neutrally stable in an unbounded fluid
\cite{Happel1983}, but that in the presence of boundaries, the
parallel orientation is more stable \cite{Huang1998}. On the other
hand, this is coherent with studies in other non-Newtonian fluids,
where ellipses settle broadside-along under gravity in Oldroyd-B
fluids \cite{Huang1998} or spherical particles aggregate
vertically during sedimentation in shear-thinning fluids
\cite{Daugan2002}. Actually, the stable orientation of long
objects under flow is determined by a competition between inertia
and viscoelasticity \cite{Liu1993}, which have opposite effects.

Fig. \ref{ForcesCoupleExpTh}a shows that the lift is oriented
downwards, as for a cambered airfoil, probably  due to the
positivity of the first normal stress difference \cite{Dollet}:
this should therefore be valid for every viscoelastic fluid
\cite{Huang1998}.

It is worth noting that lift  (Fig. \ref{ForcesCoupleExpTh}a) and
torque (Fig. \ref{ForcesCoupleExpTh}b) are not maximal at the
angle of 45$^\circ$. This contrasts with the existing prediction
of the torque exerted on an ellipse by a second-order fluid in
potential flow \cite{Wang2004}, which predicts an angular
dependence of the form $\cos\theta\sin\theta$.  We suggest two
possible explanations for this discrepancy. First, the flow of
foam is not potential, and even breaks the $x \to -x$ symmetry
between upstream and downstream \cite{DolletThese}. Second,
second-order fluids might not be good models for foams, because
they do not include yield stress.

In yield stress fluids,  viscoplastic models predict that the drag
on circular obstacles is proportional to the radius of the
obstacle \cite{Mitsoulis2004} as long as the yield stress remains
the preponderant contribution to the total stress. This agrees
with experiments on circles  \cite{Dollet2005}, and this is
compatible with the proportionality of the drag with the leading
length of the ellipse (Fig. \ref{F(Q)}). However, this scaling
with the leading length does not hold for a square obstacle, which
experiences a drag independent of its orientation
\cite{Dollet2005} for reasons we do not understand yet. In
addition, any orientation of a square obstacle is neutrally stable
in a flowing foam \cite{Dollet2005}, whereas it would align its
diagonals streamwise and spanwise in a viscoelastic liquid, as
reported in \cite{Huang1998}.

To summarise, we are not aware of a single  macroscopic, continuous
(viscoelastic or viscoplastic) model which can explain the
whole set of experimental  data.

We now propose an elastic, microscopic model, to catch the main
qualitative features of drag, lift and torque. We estimate the
contribution $\vec{F}$ of the soap film tension (which determines
the normal tensile elastic stress \cite{Weaire1999,Janiaud2005})
to the force on the ellipse. Since the foam is quasi-2D, each film
separating two bubbles in contact with the obstacle exerts on it a
force directed along the film; its magnitude is the line tension,
$\lambda$, which is twice the air-water surface tension $\gamma$,
multiplied by the foam height $h$, and a prefactor accounting for
3D geometry \cite{Dollet}. If the flow is quasistatic, the film is
along the normal $\vec{n}$ to the surface of the ellipse (see Fig.
\ref{Ellipses}). The total force is thus a sum performed over the
films in contact with the ellipse: $\vec{F} = \sum
\lambda\vec{n}$. We do not model the contribution of the bubbles
pressure, which is of the same order of magnitude, and is roughly
proportional, to the contribution of the film tension
\cite{Cox2006,DolletThese}. We do not model either the
velocity-dependent forces and torque, originating from the viscous
friction within the lubrication films between the ellipse and the
surrounding bubbles.

 If the ellipse is much larger than the bubbles, we consider the
 distance between consecutive films along the ellipse as the
 continuous function $f(\alpha)$, $\alpha$ being the
angle in the ellipse's parametric equation: $X(\alpha) =
a\cos\alpha$, $Y(\alpha) = b\sin\alpha$, and write the force
$\vec{F}$ and torque $C$ as integrals:
\begin{eqnarray}
\frac{\vec{F}}{\lambda} &=& \int \frac{\vec{n}(\alpha)}{f(\alpha)}
\mathrm{d}\alpha = ab \int_0^{2\pi} \left( \frac{\cos \alpha}{a}
\vec{e}_X + \frac{\sin \alpha}{b} \vec{e}_Y \right)
\frac{\mathrm{d}\alpha}{f(\alpha)}
, \nonumber \\
\frac{C}{\lambda} &=&
 \int \vec{r}(\alpha) \wedge \frac{\vec{n}(\alpha)}{f(\alpha)}
\mathrm{d}\alpha \cdot \vec{e}_z
  \nonumber \\
&=& (a^2 - b^2) \int_0^{2\pi} \frac{\cos \alpha \sin
\alpha}{f(\alpha)}\mathrm{d}\alpha .
  \label{CoupleIntegral}
\end{eqnarray}
   We then deduce the drag and lift as $F_x = F_X\cos\theta -
F_Y\sin\theta$, and $F_y = F_X\sin\theta + F_Y\cos\theta$,
respectively.

We must now model the function $f$, or equivalently, the
deformation of bubbles around the obstacle. As already mentioned
in \cite{Dollet}, this is strongly correlated to the local
structure of the flow: if it converges towards the obstacle
(leading side), it squashes the bubbles in contact, and $f$ is
high. Conversely, if the flow diverges from the obstacle (trailing
side), it stretches the bubbles in contact, and $f$ is low.
Experimental images support this argument (Fig. \ref{Ellipses}),
and, more precisely, lead us to set a phenomenological expression
for $f$. Fig. \ref{Ellipses} shows that the bubbles remain
squashed over the whole leading side ($\beta \leq \alpha \leq
\pi+\beta$ with $\beta = \arctan (b\cot\theta/a)$ from elementary
geometry); we thus assume that $f$ takes a maximum value, $f_M$
over this interval. At the trailing side, Fig. \ref{Ellipses}
shows that the bubbles are progressively stretched up to a maximum
close to the $y=0$ point, which is close to the angle $\alpha =
-\theta$ for simplicity. To reproduce this observation, we assume
a piecewise affine variation of $f$ from $f_M$ to a minimal value
$f_m$ in the ranges $-\theta \leq \alpha \leq \beta$, and
$\beta-\pi \leq \alpha \leq -\theta$. The analysis of several
images of the bubbles along the obstacle yields the following
estimates: $f_m = 3.3$ mm, and $f_M = 4.9$ mm. Given the aspect
ratio $a/b=1.6$, we can calculate the drag, lift and torque from
Eq. (\ref{CoupleIntegral})
  (Fig. \ref{ForcesCoupleExpTh}).


For the drag, it turns out that the result from Eq.
(\ref{CoupleIntegral}) is indiscernible (with 1\% precision, up to
a free prefactor) with Eq. (\ref{LeadingLength}); the agreement
with the experimental data is thus quite good (Fig.
\ref{ForcesCoupleExpTh}a). For the lift, we predict the sign, {\it
i.e.} explains the downwards lift: the tensile stress is larger at
the
  trailing edge where it contributes in average downwards (and
  downstream) for angles between 0$^\circ$ and 90$^\circ$, than at the
leading edge, where it contributes upwards (and upstream). This
confirms that the lift is dominated by the elasticity, as is the
case for an airfoil \cite{Dollet}. Moreover, we predict correctly
the angular dependence of the lift, and a maximum at angle
40$^\circ$, which agrees quantitatively with the experiments. For
the torque, the agreement is qualitative: we predict its sign, the
existence of  a maximum at an angle smaller than 45$^\circ$, and
the stability (instability) of the parallel (perpendicular)
orientations.

The present model relies mainly on the coupling between bubble
deformation and flow. This argument has a very general validity:
it explains the anti-inertial lift exerted by a flowing foam on an
airfoil \cite{Dollet}, and predicts quantitatively the drag on a
circle on several decades of fluid fractions \cite{Raufaste}. It
also applies in 3D, as shown by the analogies between the 2D flow
around a circle \cite{Dollet2005} and the 3D flow around a sphere
\cite{deBruyn2004,Cantat2005b}. It is qualitatively insensitive to
the presence of channel walls, both because this does not
influence the convergence or divergence of flow close to the
obstacle, and because of the very limited lateral extent of the
influence of an obstacle for foams \cite{Dollet2005,deBruyn2004}
compared to Newtonian fluids.

Foams are often modelled as viscoplastic fluids, such as Bingham
or Herschel-Bulkley models \cite{Lauridsen2002}. Such models
describe yielding, which is occurring at the leading side, where
the roughly constant amplitude of bubble deformation (Fig.
\ref{Ellipses}) is a manifestation of yield strain. On the other
hand, viscoelastic fluids such as the Oldroyd-B model often used
for polymers \cite{Larson1999} describe the delayed, elastic
response of the bubbles, apparent at the trailing side through the
progressive stretching of the bubbles (Fig. \ref{Ellipses}).

Our phenomenological model captures both the coupling between
strain and flow (with delayed response) and the saturation of
deformation (yielding). It yields a good agreement with
experimental data (Fig. \ref{ForcesCoupleExpTh}). Still, we can
suggest improvements in three directions. First, the law assumed
for $f$ is a phenomenological description of observations. The
next step would consist in predicting this function. This would
require to quantify accurately the evolution of strain due to
advection and plastic rearrangements of bubbles, which is
predicted by recent models in the simple case of shear flow
\cite{Weaire}; here, a generalization to more complex flows and
geometries is required. Second, we could extend this model to
describe the velocity-dependent contribution to drag, lift and
torque. This requires to quantify precisely the influence of
friction on the interaction between bubbles and obstacle
boundaries \cite{Denkov2005}. Third, it would be useful to include
the effect of sharp angles, in order to understand why the drag on
a square does not depend on its orientation.

%


We have benefitted from stimulating discussions with J. Wang, S. Cox and
C. Raufaste, as well as during  the FRIT workshop.

\end{document}